\documentclass[useAMS,usenatbib]{mn2e}

\usepackage{epsf}

\title[Scaling behaviour of a scalar field model of dark matter halos ]
{Scaling behaviour of a scalar field model of  \\dark matter halos }
\author[B. Fuchs and E. W. Mielke]{B. Fuchs$^{1}$\thanks{E-mail:
fuchs@ari.uni-heidelberg.de} and E. W. Mielke$^{2}$\thanks{E-mail: 
ekke@xanum.uam.mx}\\ $^{1}$Astronomisches Rechen--Institut,
M\"onchhofstr.~12--14, 69120 Heidelberg, Germany, \\
$^{2}$Departamento de F\'{\i}sica,
         Universidad Aut\'onoma Metropolitana--Iztapalapa,
       Apartado Postal 55-534, C.P. 09340, M\'exico, D.F., M\'exico}
\begin{document}

\date{Accepted . Received ; in original form 2003 September }

\pagerange{\pageref{firstpage}--\pageref{lastpage}} \pubyear{2003}

\maketitle

\label{firstpage}

\begin{abstract}
Galactic dark matter is modelled by a scalar field. In particular, it is shown
that an analytically solvable toy model with a non--linear self--interaction
potential $U(\Phi)$ leads to dark halo models which have the form of
quasi--isothermal spheres. We argue that these fit better the observed
rotation curves of galaxies than the centrally cusped halos of standard cold 
dark matter. The scalar field model predicts a proportionality between the
central densities of the dark halos and the inverse of their core radii. We 
test this prediction successfully against a set of rotation curves of low 
surface brightness galaxies and nearby bright galaxies. 
\end{abstract}


\section{Introduction}

The evidence for dark matter is overwhelming, although its nature is 
not clear. Most promising seems at present the concept of cold 
dark matter. However, this is flawed on galactic scales. One of its major 
difficulties is that models of cold dark matter halos of galaxies show inner 
density cusps, $\rho \propto r^{-\alpha}$, with $\alpha $ in the range 0.5 to
1.5 (Navarro, Frenk \& White 1997, hereafter referred to as NFW, 
Moore et al.~1999), which
are not consistent with observed rotation curves of galaxies. For this and 
other reasons alternative candidates for dark matter have been proposed, 
most prominently self--interacting cold dark mater (Spergel \& Steinhardt 2000).
Others have been based, as for example by Goodman (2000), on field theoretical 
concepts. In the same vein, Schunck (1997, 1999), Mielke \& Schunck (2002) and
Mielke et al.~(2002, hereafter referred to as MSP) have
proposed a primordial almost massless scalar field with a repulsive $\Phi^6$
self--interaction as dark matter candidate. They showed that the corresponding
non--linear Klein--Gordon equation permits a soliton solution, which is 
equivalent to a Newtonian mass configuration with the density profile of a 
quasi-isothermal sphere. Such density profiles of dark halos with homogeneous 
cores fit observed rotation curves of galaxies (de Blok, McGaugh, \& Rubin 2001,
de Blok \& Bosma 2002, hereafter referred to as dBMGR\&dBB) much better than 
cusped density profiles. Besides this interesting density profile, MSP's model
predicts a scaling law
between the central densities of the dark halos and their core radii which we
investigate here. For this purpose we recapitulate briefly the field theoretical
approach in the next section and discuss then the predicted scaling law using
decompositions of the rotation curves of a set of nearby bright and low 
surface brightness galaxies.

\section{Scalar field with a $\Phi^6$ self--interaction potential}

MSP chose ad hoc for the scalar field an analytically solvable toy model 
(Mielke 1978)
with a $\Phi^6$ type self--interaction potential,
\begin{equation}
U(\vert  \Phi \vert ) = m^2 \vert  \Phi \vert ^2 \left(1 -
\chi \vert  \Phi \vert^4 \right), \qquad \chi \vert  \Phi \vert^4\leq 1\, ,
\end{equation}
where $m$ is a tiny `bare' mass of the scalar field  and $\chi$ a coupling 
constant. Both characterize the hypothetical scalar field and are thought of as
constants of nature.
The self--interaction in the radial 
Klein--Gordon equation takes the form
$ dU(P)/dP^2=  m^2  - 3 m^2 \chi P^4$, where $P = \Phi e^{imt}$.
For a spherically symmetric configuration the corresponding
non--linear Klein--Gordon equation simplifies to an Emden type equation
\begin{equation}
P^{\prime\prime} + \frac{2}{x} P^{\prime}
  + 3\chi P^5 = 0\, ,
\end{equation}
familiar from the theory of gaseous spheres. It has 
the completely regular {\em exact} solution
\begin{equation}
P(r) = \pm \chi^{-1/4} \sqrt{\frac{A }{1+A^2 x^2}} \, , 
\end{equation}
where the dimensionless radial coordinate $x=m r$ has been introduced and 
$A = \sqrt{\chi} P^2(0)$ is related to the central value. The solution depends 
essentially on the non--linear coupling parameter $\chi$,
since the limit $\chi\rightarrow 0$ would be singular. This feature is  rather
characteristic for soliton solutions. In the following we restrict ourselves 
to the range $A\leq 1$ for which the potential $U(\vert  \Phi \vert)$ remains
positive.
 
The canonical energy--momentum tensor of a relativistic spherically symmetric 
scalar field is diagonal,
i.e.~$T_\mu{} ^\nu(\Phi) = {\rm diag} \; (\rho , -p_r,$
$-p_\bot, -p_\bot )$ with 
\begin{eqnarray}
\rho &=& \frac{1}{2} \left ( m^2 P^2  +P'^2  + U
 \right ) \, , \nonumber  \\
p_{\rm r} &=&  \rho -  U \, , \;   \\
p_\bot &=&  p_r -  P'^2 \nonumber \, ,    
\end{eqnarray}
where the prime indicates a radial derivative.
The  form (4) is familiar from perfect fluids, except that the radial and 
tangential pressures generated by the scalar field are in general different,
i.e.~$p_r \neq p_\bot $. The scalar field proposed here is not interacting by
self--gravity but exerts a gravitational force. From (4) MSP find in flat 
spacetime the energy--density 
\begin{eqnarray}
\rho &=& \frac{m^2}{2} \left [ 2 P^2
 + P'^2 - \chi P^6 \right ]  \nonumber \\
 & = & \frac{A m^2}{\sqrt{\chi}(1+A ^2 x^2)}\left[1
 +\frac{ A ^4 x^2 -A ^2}{2(1+A ^2 x^2)^2}  \right ]  \, .
\end{eqnarray}
The leading term of the Newtonian type mass concentration (5) is exactly the
density law of the quasi--isothermal sphere
\begin{equation}
\rho (r) \simeq
\frac{\rho_0 r_{\rm c}^2}{r_{\rm c}^2+r^2}  \; .
\end{equation}
At large radii the density falls of like $\rho \propto r^{-2}$ which
corresponds to an asymptotically flat rotation curve. Comparing with the MSP
model, the central density of the quasi--isothermal sphere is given by
$\rho_0 \simeq A  m^2/\sqrt{\chi}$ and the core radius is $r_{\rm c}\simeq 
1/m A$. This implies a scaling law for the dark halos of the form
\begin{equation}
\rho_0 \simeq \frac{m}{\sqrt{\chi}}\frac{1}{r_{\rm c}}\propto \frac{1}
{r_{\rm c}} \; ,
\end{equation}
where $A$, which may vary from halo to halo, {\em cancels out}.

Finally we note that this  non--linearly coupled scalar field exerts the
radial and tangential pressures
\begin{eqnarray}
p_{\rm r} &=& \frac{m^2}{2} \left [
 \chi P^6 + P'^2  \right ] =
\frac{A ^3 m^2}{2\sqrt{\chi}(1+A ^2 x^2)^2} \simeq
  \frac{A ^3 m^2}{2\sqrt{\chi}} , \label{radp} \nonumber \\
  p_\bot  &=& \frac{m^2}{2} \left [
 \chi P^6 - P'^2  \right ] =
\frac{A ^3 m^2(1-A ^2 x^2)}{2\sqrt{\chi}(1+A ^2 x^2)^3} \simeq
  \frac{A ^3 m^2}{2\sqrt{\chi}} ,
\end{eqnarray}
respectively.
Thus, at the center the pressure is isotropic, $p_{\rm r}(0)= p_\bot(0)$, 
whereas asymptotically at infinite radius
\begin{equation}
p_{\rm r}\, , - p_\bot \rightarrow \frac{m^2}{2\sqrt{\chi} A  x^4} \; .
 \label{radpa}
\end{equation}

\section{Astronomical tests}

We have tested the theoretical model predictions against rotation curve data of
a set of low surface brightness galaxies taken from dBMGR\&dBB. 
The authors have measured 
high--resolution rotation curves of in total 54 galaxies. For about half of 
them surface photometry is available. For these galaxies the authors do not
provide only kinematical data, but have also constructed dynamical models of 
the galaxies. The observed rotation curves are modeled as
\begin{equation}
  v_{\rm c}^2(R)= v_{\rm c,bulge}^2(R)+v_{\rm c,disc}^2(R)+
  v_{\rm c,isgas}^2(R)+v_{\rm c,halo}^2(R),
\end{equation}
where $v_{\rm c,bulge}$, $v_{\rm c,disc}$, $v_{\rm c,isgas}$, and 
$v_{\rm c,halo}$ denote the contributions due to the bulge, the stellar disc, 
the interstellar gas, and the dark halo, respectively. The radial variations of
$v_{\rm c,bulge}(R)$, $v_{\rm c,disc}(R)$, and $v_{\rm c,isgas}(R)$ were derived
from the observations, while the normalizations by the mass--to--light ratios
were left as free parameters of the fits of the mass models to the data. Fits of
the form (10) to observed rotation are notoriously ambiguous. Thus, dBMGR\&dBB
provide for each galaxy several models, one with zero bulge and 
disc mass, one model with a `reasonable' mass--to--light ratio of the bulge and
the disc, and finally a `maximum--disc' model with bulge and disc masses at the
maximum allowed by the data. Furthermore these authors try for each galaxy
two types of dark halo models. One is the cusped NFW density law 
and the second is the quasi--isothermal sphere. While varying
the disc contribution to the observed rotation curve leads to fits of the same
quality, dBMGR\&dBB find that
the quasi--isothermal sphere models of the dark halos give significantly better
fits to the data than the cusped NFW density law. Thus the scalar field model 
presented here is in this aspect even superior to the cold dark matter model 
in its present form. 

Prada et al.~(2003) have attempted using SDSS data on
satellite galaxies of isolated host galaxies to probe on 100 kpc scale the {\em
outer} halo mass distributions. They find that the line--of--sight velocity
dispersions of the satellites follow closely the radially declining velocity
dispersion profile of halo particles in a NFW halo. This implies an outer mass
density distribution of the form $\rho \propto r^{-3}$ which is at variance with
the prediction of the quasi--isothermal sphere (6). In the cold dark matter
model the system of satellite galaxies is assembled during the same accretion
processes as the dark halo, and Prada et al.~(2003) assume consistently for the
satellites the same distribution function in phase space as for the halo
particles. In the scalar field model, however, the dark halo provides for the
baryonic matter simply a Newtonian force field. The distribution function of
satellite galaxies in phase space is thus not specified and can be modelled
according to the observations, even if the potential trough of the
quasi--isothermal sphere has a shallower profile than a NFW halo. 

\begin{figure}
\begin{center}
\epsfxsize=8.0cm
   \leavevmode
      \epsffile{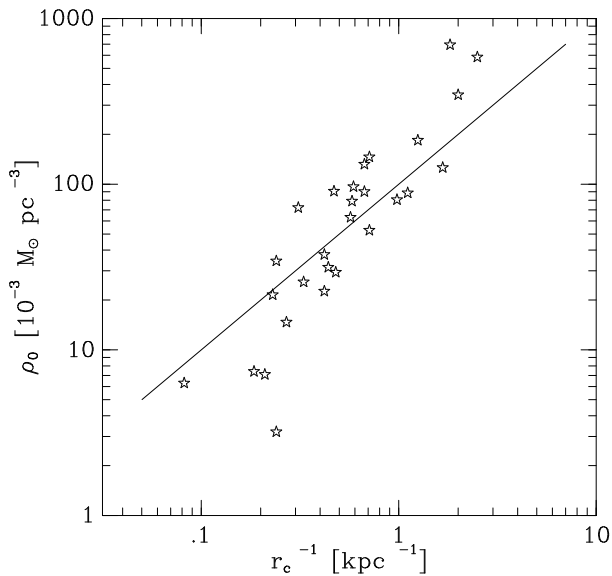}
 \caption{Central densities versus the inverses of the core radii of dark halos
  of low surface brightness galaxies derived from modelling their rotation 
  curves. The halo models are constructed assuming `realistic' mass--to--light
  ratios for the discs. The solid line is the predicted  $\rho_0 \propto 
  r_{\rm c}^{-1}$ relation.}
         \label{imag1}
   \end{center}
   \end{figure}
   \begin{figure}
\begin{center}
\epsfxsize=8.0cm
   \leavevmode
      \epsffile{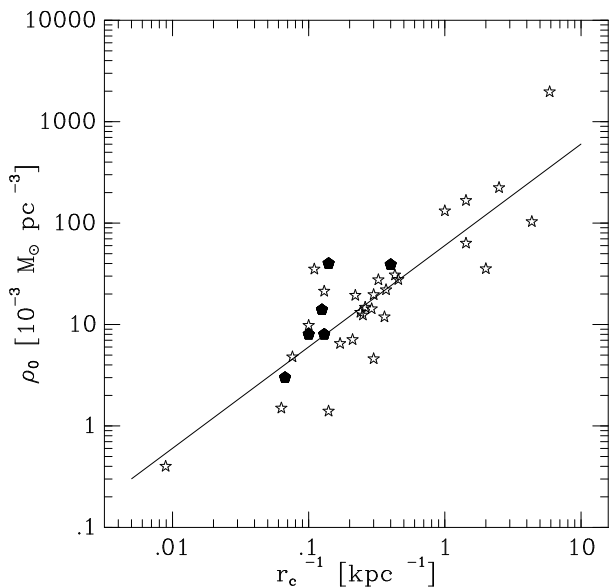}
 \caption{Same as Fig.~1, but the halo models are constructed assuming `maximum 
  discs'. Open symbols: low surface brightness galaxies, filled symbols: nearby
  bright galaxies}
         \label{imag2}
   \end{center}
   \end{figure}
Next we examine the predicted scaling relation (7). A relation of this type was
found empirically by Salucci \& Burkert (2000), although this was based on the
universal rotation curve model of Persic, Salucci \& Stel (1996) and the,
also empirically derived, halo density profile of Burkert (1995). This density 
law resembles the quasi--isothermal sphere in that it has also a homogeneous 
core. In Fig.~1 we show central densities versus the inverses of the core radii
of quasi--isothermal dark halo models constructed by dBMGR\&dBB assuming for
the discs mass--to--light ratios consistent with current population synthesis 
models. There is despite some scatter a clear correlation between $\rho_0$ and 
$r_{\rm c}^{-1}$ over several orders of magnitude, $\log(\rho_0) \propto 
(1.46\pm 0.55)\log(r_{\rm c}^{-1})$. Thus the dark halo model data seem to 
confirm statistically the scaling 
relation (7). The scatter in the correlation diagram is probably due to the 
near degeneracy of the fits to the observed rotation curves, even if the disc 
contributions to the rotation curves are fixed. Moreover practice shows that 
the radial exponential scale lengths of the discs cannot be determined more
precisely than about 20\%. This has a considerable effect for the disc 
contributions to the rotation curves (cf.~equation 10) and consequently for the
dark halo models and might also explain some of the scatter of the 
correlation diagram in Fig.~1. We believe, however, that this has not changed
the general trend. Using arguments of the density wave theory of galactic 
spiral structure, one of us has pointed out, judging from the implied internal 
dynamics of the galactic discs, that the discs might be near to maximum 
(Fuchs 2002, 2003a, b). This would imply for the discs of some galaxies 
mass--to--light 
ratios which are significantly higher than in current population synthesis
models of the discs of low surface brightness galaxies. These can be
modified, though, to yield higher mass--to--light ratios (Lee et al.~, in 
preparation). Therefore we show in Fig.~2 the central densities 
versus the inverses of the core radii of the dark halo models, if they are 
constructed assuming `maximum discs'. Although the dark halo parameters are 
shifted to other values, the linear correlation persists. With $\log(\rho_0) 
\propto (1.08 \pm 0.39)\log( r_{\rm c}^{-1})$ the dark halo models fit nearly
ideally to the predicted 
scaling relation (7). Included into Fig.~2 are also `maximum disc' dark halo 
parameters of nearby bright galaxies (Fuchs  1999, 2003a, Fried \& Fuchs, in 
preparation) which fit also well to the scaling law. We conclude from this 
discussion that the astronomical tests are rather encouraging for the scalar 
field model of dark halos presented here. What is missing at present, is of 
course a cosmological setting of the scalar field model. We hope to address
this and, in particular, the question of large scale structure formation, which
is on the other hand successfully described by cold dark matter cosmology, in 
the future.

\section*{Acknowledgments}

We would like to thank the organizers of the Tenth Marcel Grossmann Meeting, 
in particular Remo Ruffini, for providing us the opportunity to discuss these 
topics in Rio de Janeiro in July 2003 and to meet again after our joint student
times in the mid 1970th at the University of Kiel.

\bsp

\label{lastpage}

\end{document}